# Sculpting of Exoplanetary Systems Driven by a Misaligned Disk and Stellar Oblateness: Origin of Perpendicular Orbits in HD 3167

Tao Fu(伏韬) ORCID:0000-0002-9981-0761, Yue Wang(王悦)[1] ORCID:0000-0001-9172-0895

## Abstract

A significant proportion of exoplanets have been detected with highly tilted or even polar orbits relative to their host stars' equatorial planes. These unusual orbital configurations are often linked to post-disk secular interactions among multiple bodies. However, many aspects remain elusive. In this study, we investigate the role of disk-induced spin-orbit misalignments in shaping architecture of multi-planet systems, taking into account the combined effect of the host star's oblateness and the full-space disk potential. We demonstrate that large mutual planetary inclinations can arise from a saddle-center bifurcation occurring during the photoevaporation of the disk. This bifurcation triggers an instant, non-adiabatic transition in the planet's libration. Following this process, the orbital evolution diverges into several distinct patterns. Notably, in scenarios involving a near-polar primordial misalignment, the orbit, consistently librating about a coplanar equilibrium axis, can be captured by an orthogonal equilibrium during the decay of the stellar oblateness. However, the orbit will be eventually recaptured by the coplanar equilibrium, aligned or anti-aligned with the orientation of the outer orbit, resulting in either a prograde or retrograde inner-outer orbit configuration. Additionally, general relativity contributes to maintaining eccentricity stability within these dynamic scenarios. Through the proposed mechanism, we can provide a plausible explanation for the unique, near-perpendicular and likely retrograde orbit architecture observed in the HD 3167 system, enhancing our understanding of exoplanetary system dynamics.

[1] Corresponding author ywang@buaa.edu.cn
School of Astronautics, Beihang University, Beijing 102206, China

## 1. Introduction

Contrary to the coplanarity observed in our solar system, spin-orbit misalignments are prevalent in exoplanetary systems. Recent Rossiter-McLaughlin (RM) effect observations suggest that approximately 40% of star hosting hot Jupiters, as well as a substantial number of smaller planets, exhibit high obliquities. These misalignments have typically been attributed to secular multi-body interactions, such Kozai cycles (Fabrycky & Tremaine 2007), planet-planet scatterings (Beaugé & Nesvorný 2012), and chaotic interactions between planets and stellar spin (Storch et al. 2014). However, none of these mechanisms alone can account for the observed obliquity distributions, given their specific triggering requirements.

An alternative explanation for such high stellar obliquities involves primordial protoplanetary disk misalignments, which can result from chaotic accretion (Thies et al. 2011; Fielding et al. 2015; Bate 2018), magnetic interactions (Foucart & Lai 2011; Lai et al. 2011), or torques from a distant binary companion (Batygin 2012; Batygin & Adams 2013; Lai 2014; Zanazzi & Lai 2018). These mechanisms can generate a broad distribution of primordial obliquities (Albrecht et al. 2022). By considering these disk-induced spin-orbit misalignments, several critical issues in exoplanetary systems can be addressed. Recent studies suggest that inclined stellar oblateness may contribute to the production of mutual inclinations between planets, offering an potential underlying mechanism for the well-known phenomenon of *Kepler* dichotomy (Spalding & Batygin 2016; Li et al. 2020; Spalding & Millholland 2020). The retrograde orbits of the two coplanar planets in the K2-290A system could plausibly be attributed to primordial disk misalignments (Hjorth et al. 2021). Furthermore, it has been shown that the combined effects of high-eccentricity migration and primordial stellar obliquity can produce a specific subset of hot Jupiters favoring perpendicular orbits, resembling the recently observed prevalence of perpendicular planets close to their host stars (Albrecht et al. 2021; Vick et al. 2023).

These studies have predominantly focused on scenarios involving either small primordial misalignments or the assumption that planets remain aligned with their respective protoplanetary disks. Typically, these studies utilize a secular disk potential approximation, assuming planets are close to their disk planes. However, we will demonstrate that when planets are displaced from the disk plane due to inclined perturbations, stemming either from an inner stellar oblateness or the gravitational influence of outer massive planets, the dynamics under a full-space disk potential lead to markedly different evolutionary behaviors. This full-space model reveals a qualitatively distinct pathway in the evolutionary dynamics of exoplanetary systems, offering a new perspective for understanding their long-term orbital behaviors.

In this work, we explore the characteristics of dynamical evolution of planets initially embedded within a highly misaligned protoplanetary disk. We find that a saddle-center bifurcation, occurring during the photoevaporation of the disk, can induce an instant, non-adiabatic transition in the planet's orbit libration, leading to large mutual planetary inclinations. This process introduces a novel aspect of planetary evolutionary dynamics. A particularly intriguing scenario involves a near-polar initial misalignment: the orbit, initially librating around the coplanar equilibrium axis, may be captured by an orthogonal equilibrium as stellar oblateness decreases. This process can ultimately result in final near-perpendicular, even retrograde inner-outer orbit configurations, resembling those observed in many exoplanetary systems (Petrovich et al. 2020).

We show that the newly proposed mechanism can be well applied to the HD 3167 system, providing a plausible explanation for its unique, perpendicular orbital configuration between the inner and outer planets. Recent observations reveal that the HD 3167 has four planets. HD 3167 b, an ultra-short-period planet (USP), has an orbit closely aligned with the stellar spin ($\psi_\star = 29.5^{+7.2^\circ}_{-9.4}$), whereas the outer planet, HD 3167 c, has an orbit nearly perpendicular to the orbit of HD 3167 b ($i_{\rm bc} = 102.3^{+7.4^\circ}_{-8.0}$ and $\psi_\star = 107.7^{+5.1^\circ}_{-4.9}$) (Bourrier et al. 2021). Dynamical analysis suggests that HD 3167 d shares a nearly coplanar orbit with HD 3167 c (Dalal et al. 2019). The peculiar architecture of HD 3167 has attracted many attentions. Dalal et al. (2019) and Bourrier et al. (2021) thought that the large obliquity of the outer planets may be attributed to an further out massive companion, similar to the framework proposed for Kepler-129 (Zhang et al. 2021). The fourth planet, HD 3167 e, has indeed been recently detected. However, it is strongly coupled and likely coplanar with HD 3167 d-c, making it an unlikely source for such a configuration. Furthermore, other potential companions that could be responsible for production of the present-day architecture of HD 3167 has been ruled out through RV and direct imaging data (Bourrier et al. 2022). Thus, the dynamical history of HD 3167 remains a mystery until now.

Below, in Sect. 2, we first explore the mutual inclination excitation process of multiple planets initially embedded within their host star's protoplanetary disk, through analyzing the equilibria and their bifurcations during photoevaporation of the inner region of the disk. In Sect.3, we show that the general eccentricity excitation and dynamical instability resulting from such a dynamical bifurcation can be tamed by general relativity. In Sect. 4, we apply the proposed mechanisms to the HD 3167 system. In Sect. 5 and Sect.6, we discuss and summarize.

## 2. Excitation of Mutual Inclinations

Planets that form within their protoplanetary disks are initially circular and well-aligned with the disk. However, as the disk disperses, out-of-plane perturbations, such as the gravitational influence of an inclined outer companion or inclined stellar oblateness, can drive the planets out of the disk plane, following perturbed *n*-body behaviors (Picogna & Marzari 2015; Martin et al. 2016; Zanazzi & Lai 2017). In this section, we consider the dynamical evolution of a two-planet system initially embedded within a misaligned disk, driven by the combined effects of the stellar oblateness and the full-space disk potential. A simplified secular model, assuming circular planetary orbits, is developed. Based on this model, the equilibria and their bifurcations are analyzed to provide analytical insights into the secular dynamics.

## 2.1 Problem Set-up

Let's start with a practical exoplanetary system, the HD 3167 system. As mentioned above, it has four planets in a special configuration. We show that the secular dynamics of this system can be significantly simplified by assuming the outer three planets, HD 3167 d, c, and e, to be effectively locked together (see Figure 1). The validity of this assumption can be evaluated based on their capacity to resist differential precessions induced by the inclined perturbations from the inner planet (HD 3167 b), which can be parameterized by (Zanazzi & Lai 2017)

$$\delta_{\mathrm{dc,b}} \simeq \frac{\Omega_{\mathrm{d,b}} - \Omega_{\mathrm{c,b}}}{\Omega_{\mathrm{d,c}} + \Omega_{\mathrm{c,d}}} \tag{1}$$

Here, $\Omega_{i,j}$ represents the precession frequency of $\boldsymbol{l}_i$ around $\boldsymbol{l}_j$ under their mutual interactions, in which $\boldsymbol{l}_i$ is the normalized angular momentum of planet $i$. IF $\delta_{\mathrm{dc,b}} \ll 1$, the assumption is feasible. Although, the inclined stellar oblateness may also disrupt the couplings among outer planets, it generally has a weaker effect than the inclined innermost planet.

The precession frequency $\Omega_{i,j}$ can be computed by

$$\Omega_{i,j} = \frac{\phi_{i,j}}{L_i} \tag{2}$$

where $\phi_{i,j}$ is the coefficient characterizing the magnitude of the interaction potential between planets $i$ and $j$, given by equation (A3) or

(A5). Then, $\delta_{\mathrm{dc,b}}$ can be reformed as

$$\delta_{\mathrm{dc,b}} = \frac{\phi_{\mathrm{b,c}}}{\phi_{\mathrm{d,c}}} \frac{L_{\mathrm{c}}/L_{\mathrm{d}}\left(\phi_{\mathrm{b,d}}/\phi_{\mathrm{b,c}}\right)-1}{1+L_{\mathrm{c}}/L_{\mathrm{d}}} \tag{3}$$

where $L_i = m_i\sqrt{Gm_\star a_i}$ is the magnitude of the angular momentum of planet $i$.

By using the parameters of HD 3167 presented in Table 1, it results in $\delta_{\mathrm{dc,b}} \approx 0.15$. Consequently, the assumption of the lock of the three outer planets is valid. This assumption is equivalent to replacing the outer three planets with a single massive planet with the orbital angular momentum $\boldsymbol{L}_{\mathrm{dce}} = L_{\mathrm{dce}}\hat{\boldsymbol{l}}_{\mathrm{dce}} = \left(L_{\mathrm{c}} + L_{\mathrm{d}} + L_{\mathrm{e}}\right)\hat{\boldsymbol{l}}_{\mathrm{dce}}$.

In addition, since $L_{\mathrm{b}}/L_{\mathrm{dec}} \approx 0.05 \ll 1$, the outer angular momentum $\boldsymbol{l}_{\mathrm{dce}}$ can be considered fixed, being able to remain the orientation of the protoplanetary disk after the disk's dissipation.

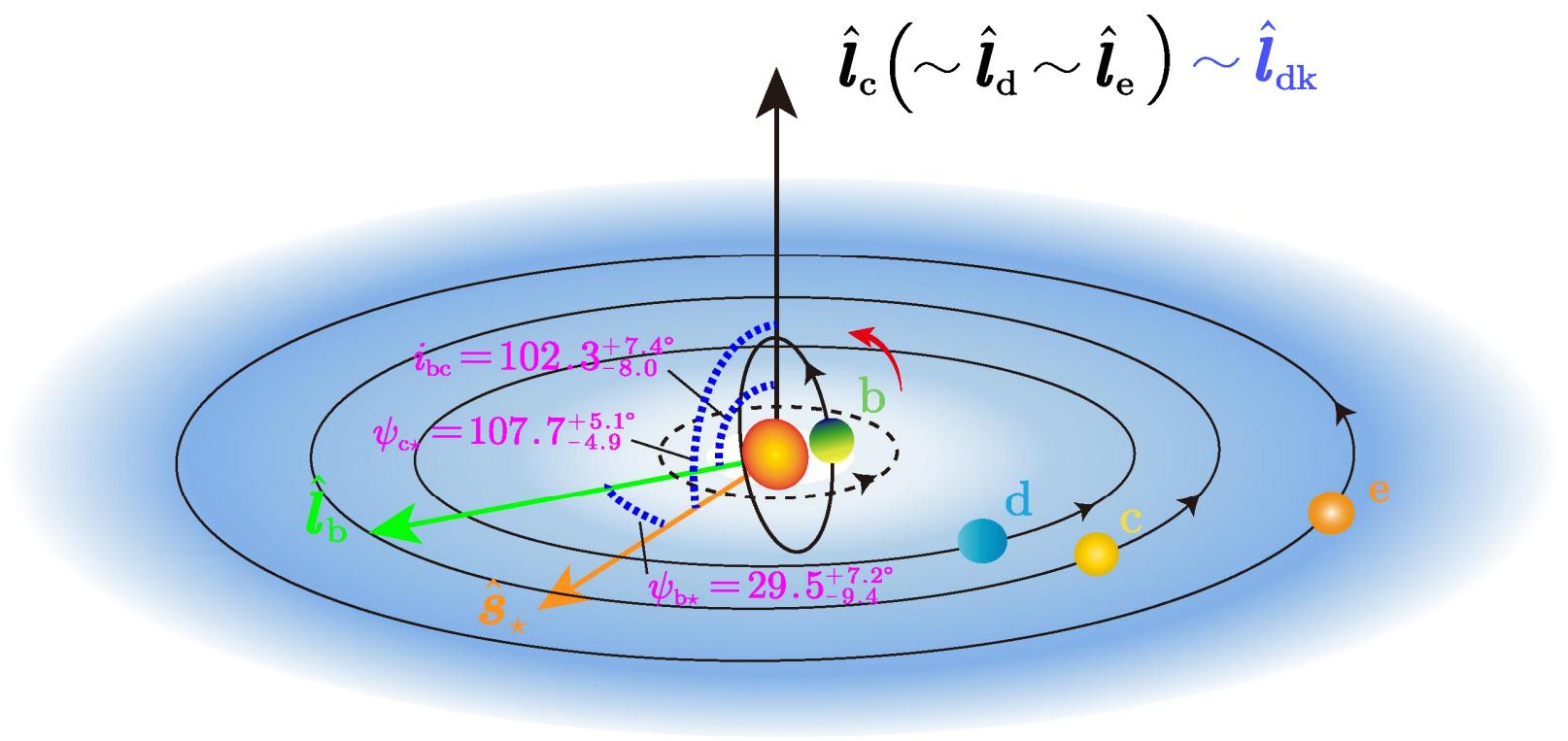

**Figure 1.** Schematic illustration of the dynamical history of the HD 3167 system (not to scale). Consider a primordial near-polar misalignment between the disk (orientated along $\hat{\boldsymbol{l}}_{\mathrm{dk}}$) and the stellar spin (orientated along $\hat{\boldsymbol{s}}_\star$). The planet b, initially aligned with the disk, transitions to a state close alignment with the stellar spin during the photoevaporation of the inner region of the disk. The outer three planets (d, c, and e) inherit the orientation of the disk $\hat{\boldsymbol{l}}_{\mathrm{dk}}$. The outer orbits are strongly coupled, allowing the assumption of their lock together (see the text). The angles indicated in the figure are the observation constrained values

of the HD 3167 system. (See Sect. 4 for details on the dynamical evolution of HD 3167.)

Consequently, in this section, we first consider the dynamics of a two-planet system initially embedded within a protoplanetary disk that is misaligned with the stellar spin axis. The two planets, denoted as planets b and c with masses $m_{\rm b}$ and $m_{\rm c}$, and semi-major axes $a_{\rm b}$ and $a_{\rm c}$ ($a_{\rm b} < a_{\rm c}$), evolve under the secular effects of stellar oblateness, full-space disk potential, and their mutual planetary interactions, as well as the photoevaporation of the disk and the decay of the stellar oblateness.

**Table 1 Parameters of HD 3167 and the Fiducial System**

| **Parameter** | **Value** |
|---|---|
| Parameters of HD 3167 | |
| $R_\star$ | $0.871 \pm 0.006 R_\odot$ |
| $m_\star$ | $0.852^{+0.026}_{-0.015}\, m_\odot$ |
| $m_{\rm b}$ | $4.73^{+0.28}_{-0.29}\, m_\oplus$ |
| $a_{\rm b}$ | $0.01802 \pm 0.00025$ au |
| $m_{\rm c}$ | $10.67^{+0.85}_{-0.81}\, m_\oplus$ |
| $a_{\rm c}$ | $10.67^{+0.85}_{-0.81}\, m_\oplus$ au |
| $m_{\rm d} \sin i_{\rm d}$ | $5.03 \pm 0.5 m_\oplus$ |
| $a_{\rm d}$ | $0.07703^{+0.00106}_{-0.00108}$ au |
| $m_{\rm e} \sin i_{\rm e}$ | $9.74^{+1.20}_{-1.15}\, m_\oplus$ |
| $a_{\rm e}$ | $0.4048^{+0.0077}_{-0.0074}$ au |
| Fiducial system parameters | |
| $m_\star$ | $1 m_\odot$ |
| $R_\star$ | $1 R_\odot$ |
| $m_{\rm b}$ | $5 m_\oplus$ |
| $a_{\rm b}$ | $0.05$ au |
| $m_{\rm c}$ | $1 m_{\rm J}$ |
| $a_{\rm c}$ | $1$ au |
| $J_{2,0}$ | $\sim 10^{-3} - 10^{-5}$ |
| $\tau_\star$ | 1 Myr |
| $m_{\rm dk,0}$ | $0.05 m_\odot$ |
| $r_{\rm c}$ | 2 au |
| $r_{\rm out}$ | 50 au |
| $\tau_{\rm v,in}$ | $1 \times 10^4$ yr |
| $\tau_{\rm v,out}$ | 1 Myr |

**References.** Bourrier et al. 2021, 2022 for HD 3167.

## 2.2 Equilibrium and Bifurcation

To obtain an analytical insight into the characteristics of the secular dynamics, in this section, we consider a simplified secular model based on the assumption that the all the orbits are circular ($e(s)=0$) and the outer orbit is fixed due to $L_{\mathrm{b}} \ll L_{\mathrm{c}}$. Then, the motion of $\boldsymbol{l}_{\mathrm{b}}$, derived by substituting the secular potential, equation (A1), into equation (A18), is given by

$$\dot{\boldsymbol{l}}_{\mathrm{b}} = -\Omega_{\mathrm{b}\star}\left(\hat{\boldsymbol{l}}_{\mathrm{b}} \cdot \hat{\boldsymbol{s}}_{\star}\right)\hat{\boldsymbol{s}}_{\star} \times \boldsymbol{l}_{\mathrm{b}} - \Omega_{\mathrm{b,c}}\left(\boldsymbol{l}_{\mathrm{b}} \cdot \hat{\boldsymbol{l}}_{\mathrm{c}}\right)\hat{\boldsymbol{l}}_{\mathrm{c}} \times \boldsymbol{l}_{\mathrm{b}} - \Omega^{*}_{\mathrm{b,dk}}\left(\hat{\boldsymbol{l}}_{\mathrm{b}} \cdot \hat{\boldsymbol{l}}_{\mathrm{dk}}\right)\hat{\boldsymbol{l}}_{\mathrm{dk}} \times \boldsymbol{l}_{\mathrm{b}} \tag{4}$$

Here, the precession frequencies $\Omega_{\mathrm{b}\star} = \phi_{\mathrm{b}\star}/L_{\mathrm{b}}$, $\Omega_{\mathrm{b,c}} = \phi_{\mathrm{bc}}/L_{\mathrm{b}}$, and

$$\Omega^{*}_{\mathrm{b,dk}} = \frac{\Omega_{\mathrm{b,dk}}}{\sin\theta_{\mathrm{b,dk}}}\left(1 - \frac{\pi}{4}\sin\theta_{\mathrm{b,dk}}\right), \tag{5}$$

where $\Omega_{\mathrm{b,dk}} = \phi_{\mathrm{dk,b}}/L_{\mathrm{b}}$ and $\theta_{\mathrm{b,dk}}$ is the angle between $\hat{\boldsymbol{l}}_{\mathrm{b}}$ and $\hat{\boldsymbol{l}}_{\mathrm{dk}}$.

It is obvious that the equation of motion remains unchanged if we perform the transformations $(\boldsymbol{l}_{\mathrm{b}}, t) \rightarrow (-\boldsymbol{l}_{\mathrm{b}}, -t)$ or $\boldsymbol{e} \rightarrow -\boldsymbol{e}$ or $\hat{\boldsymbol{s}}_{\star} \rightarrow -\hat{\boldsymbol{s}}_{\star}$ or $\hat{\boldsymbol{l}}_{\mathrm{c}} \rightarrow -\hat{\boldsymbol{l}}_{\mathrm{c}}$ or $\hat{\boldsymbol{l}}_{\mathrm{dk}} \rightarrow -\hat{\boldsymbol{l}}_{\mathrm{dk}}$. The symmetries indicated by these transformation invariances can be utilized to simplify the explorations of the secular dynamics (Wang & Fu 2023).

By using equation (4), the equilibria of $\boldsymbol{l}_{\mathrm{b}}$, at which the orbit will remain at rest, can be solved from $\dot{\boldsymbol{l}}_{\mathrm{b}} = 0$,

$$\begin{aligned} &\left(\boldsymbol{l}_{\mathrm{b}} \cdot \hat{\boldsymbol{s}}_{\star}\right)\hat{\boldsymbol{s}}_{\star} \times \boldsymbol{l}_{\mathrm{b}} + \left\{\frac{1/\varepsilon_{\star\mathrm{dk}}}{\sin\theta_{\mathrm{b,dk}}}\left(1 - \frac{\pi}{4}\sin\theta_{\mathrm{b,dk}}\right) + 1/\varepsilon_{\star\mathrm{c}}\right\}\left(\boldsymbol{l}_{\mathrm{b}} \cdot \hat{\boldsymbol{l}}_{\mathrm{dk}}\right)\hat{\boldsymbol{l}}_{\mathrm{dk}} \times \boldsymbol{l}_{\mathrm{b}} \\ &\triangleq \left(\boldsymbol{l}_{\mathrm{b}} \cdot \hat{\boldsymbol{s}}_{\star}\right)\hat{\boldsymbol{s}}_{\star} \times \boldsymbol{l}_{\mathrm{b}} + \left\{1/\varepsilon^{*}_{\star\mathrm{dk}} + 1/\varepsilon_{\star\mathrm{c}}\right\}\left(\boldsymbol{l}_{\mathrm{b}} \cdot \hat{\boldsymbol{l}}_{\mathrm{dk}}\right)\hat{\boldsymbol{l}}_{\mathrm{dk}} \times \boldsymbol{l}_{\mathrm{b}} = 0 \end{aligned} \tag{6}$$

where, we assume the outer orbit is aligned with the disk, i.e., $\hat{\boldsymbol{l}}_{\mathrm{dce}} = \hat{\boldsymbol{l}}_{\mathrm{dk}}$, and the two dimensionless parameters $\varepsilon_{\star\mathrm{c}}$ and $\varepsilon_{\star\mathrm{dk}}$ are defined by

$$\varepsilon_{\star c} = \frac{\Omega_{b\star}}{\Omega_{b,c}}, \tag{7}$$

and

$$\varepsilon_{\star dk} = \frac{\Omega_{b\star}}{\Omega_{b,dk}}, \tag{8}$$

which characterize the magnitudes of the perturbations from planet c and the disk gravity relative to the stellar oblateness, respectively.

For equation (6), it can be obtained that either $\boldsymbol{l}_b \cdot \hat{\boldsymbol{s}}_\star = \boldsymbol{l}_b \cdot \hat{\boldsymbol{l}}_{dk} = 0$ or $\boldsymbol{l}_b \cdot \left(\hat{\boldsymbol{s}}_\star \times \hat{\boldsymbol{l}}_{dk}\right) = 0$ by taking scalar product of this equation with $\hat{\boldsymbol{s}}_\star$ and with $\hat{\boldsymbol{l}}_{dk}$, respectively. In the first case, the solution of $\boldsymbol{l}_b$ is perpendicular to the plane containing $\hat{\boldsymbol{s}}_\star$ and $\hat{\boldsymbol{l}}_{dk}$, referred as the orthogonal circular equilibrium; while in the second case, $\boldsymbol{l}_b$ lies in this plane, referred as the coplanar circular equilibrium (notations refer to Tremaine et al. (2009)).

In the case of coplanar equilibria, the direction of $\boldsymbol{l}_b$ can be specified by its azimuthal angle $\bar{\psi}_b$ relative to the stellar spin $\hat{\boldsymbol{s}}_\star$, and the angle relative to the disk is thereby $\psi_{\star 0} - \bar{\psi}_b$, where $\psi_{\star 0}$ is the angle between the disk and stellar spin (i.e., the primordial obliquity). Note that the angle $\psi_{\star 0} - \bar{\psi}_b$ sightly differs from $\theta_{b,\,dk}$ used in equation (5), because $\theta_{b,\,dk}$ is defined in the range $[0,\, 180°]$, while $\psi_{\star 0} - \bar{\psi}_b$ varies from 0 to $360°$. Then, equation (6) becomes

$$\sin 2\bar{\psi}_b - \left(1/\varepsilon^*_{\star dk} + 1/\varepsilon_{\star c}\right)\sin 2\left(\psi_{\star 0} - \bar{\psi}_b\right) = 0\ . \tag{9}$$

At first glance, it appears that if $\bar{\psi}_b$ is a solution, then $\bar{\psi}_b \pm \pi/2$ and $\bar{\psi}_b + \pi$ are also solutions, as the cases of the classical Laplace surface (Tremaine et al. 2009; Saillenfest & Lari 2021). However, in our problem, $\varepsilon^*_{\star dk}$ is also a function of $\bar{\psi}_b$ in the form of $\sin\theta_{b,\,dk} = \left|\sin\left(\psi_{\star 0} - \bar{\psi}_b\right)\right|$, as indicated by equation (5). This implies that if $\bar{\psi}_{\rm b}$ is indeed a solution, $\bar{\psi}_{\rm b} \pm \pi/2$ are not solutions, except when $1/\varepsilon^{*}_{\star \rm dk}$ becomes negligible compared to $1/\varepsilon_{\star \rm c}$. However, $\bar{\psi}_{\rm b} + \pi$ is still a solution. Therefore, we only need to focus on solutions located in the range $[0, 180°)$.

Obtaining analytical solutions for equation (9) proves to be challenging. Figure 2 presents the numerical solutions over time for a fiducial two-planet system. As depicted in Figure 2A, the coplanar equilibria evolve and bifurcate in response to the changes in the parameters $\varepsilon_{\star \rm dk}$ and $\varepsilon_{\star \rm c}$, due to the decay of the disk and the stellar oblateness. In Figure 2B, the phase portraits of the system are illustrated.

The evolution of the system can be divided into four distinct phases based on the geometry of the phase portraits. In the initial phase, $\varepsilon_{\star \rm dk} \ll 1$ and $\varepsilon_{\star \rm c} \gg 1$, the planetary motion is dominated by the disk potential. This phase features two coplanar equilibria, denoted as $L_2$ and $L_3$, and one orthogonal equilibrium, denoted as $L_1$. The phase portrait in this phase, shown in panel (a) of Figure 2B, resembles the classical Laplace surface (Saillenfest & Lari 2021; Tremaine et al. 2009). The trajectories passing through $L_2$ separates the stable libration regions around $L_1$ and $L_3$.

As the dissipation of the disk, the parameter $\varepsilon_{\star \rm dk}$ increases, and when it reaches about $0.55$, a bifurcation point referred as $B_1$ arises, where $\varepsilon_{\star \rm c}$ remains $\gg 1$. At this point, one equilibrium bifurcates into two new equilibria, a center (denoted as $P_5$) and a saddle point (denoted as $P_4$), as shown by the panel (b) of Figure 2B. As a result, in the second phase (post-$B_1$), the system features five equilibria, three centers and two saddle points. The equilibrium $L_3$ closely aligns with the disk orientation, while $P_5$ approximately aligns with the stellar spin axis.

The trajectories passing through $P_4$ serve as the separatrix.

Subsequently, as the continuous increases of $\varepsilon_{\star\mathrm{dk}}$, the saddle point $P_4$ gradually approaches the equilibrium $L_3$, as shown in the lower panel of Figure 2A. This process culminates in the merging of the two equilibria at bifurcation point $B_2$, where $\varepsilon_{\star\mathrm{dk}} = 3.3$ and $\varepsilon_{\star\mathrm{c}} \gg 1$. Then, in the third phase (post-$B_2$), the equilibria $P_5$ and $L_1$ with the separatrix emerging from $L_2$ determines the structure of the phase space, as illustrated in panel (c) of Figure 2B.

During the third phase, initially, the equilibrium $P_5$ is closely aligned with the stellar spin. However, as the stellar oblateness decays, $P_5$ gradually shifts towards the orientation of the outer planet. Eventually, when the stellar oblateness weakens sufficiently ($\varepsilon_{\star\mathrm{c}} \ll 1$), the equilibrium within the range of $0-90°$ realigns with the orientation of the outer orbit. The phase portrait of the system in this phase is depicted in panel (d) of Figure 2B.

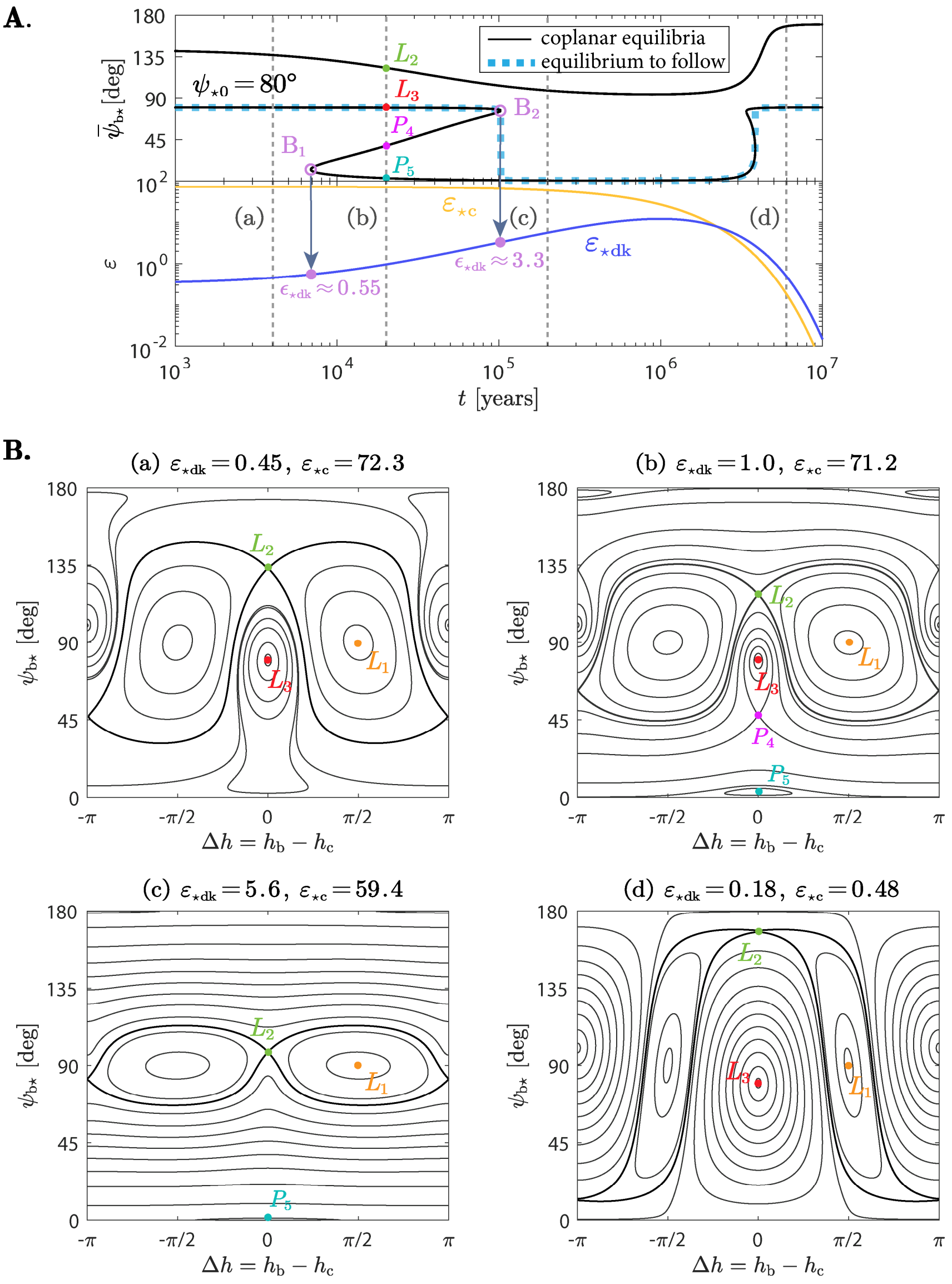

**Figure 2.** Equilibria and bifurcations of a fiducial two-planet system. The results are obtained for our fiducial system by using $J_{2,0} = 5\times10^{-4}$, $\psi_{\star 0} = 80°$, and other parameters are those listed in Table 1. Figure 2A shows all coplanar equilibrium solutions within the 0–180° range. The solutions evolve over time as shown in the upper panel, where the sky-blue dashed curve represents the equilibrium path that a planet initially aligned with the disk is likely to follow. The lower panel tracks the evolution of the parameters $\varepsilon_{\star dk}$ and $\varepsilon_{\star c}$. Figure 2B illustrates the phase portraits of the system by plotting the level curves of the integral of motion $\Phi$ given by equation (A1) in the $(\Delta h, \psi_{b\star})$ space, where $\Delta h = h_b - h_c$ represents

the differences in the longitude of ascending node of the inner and outer orbits. Panels (a)-(d) correspond to the four stages during the evolution highlighted in Figure 2A. Note: $\Delta h = 0$ or $\pi$ denotes the alignment or anti-alignment of $\boldsymbol{l}_{\rm b}$ with $\boldsymbol{l}_{\rm c}$ (or $\boldsymbol{l}_{\rm dk}$), and $\Delta h = \pm\pi/2$ denotes the perpendicularity between $\boldsymbol{l}_{\rm b}$ and $\boldsymbol{l}_{\rm c}$ (or $\boldsymbol{l}_{\rm dk}$).

From the fiducial example, we demonstrate that the bifurcations $B_1$ and $B_2$, which occur during the photoevaporation of the disk, significantly influence the geometry of the system's phase portraits. In this example, the parameter $\varepsilon_{\star{\rm c}} \gg 1$ and remains nearly unchanged during the first three phases. Therefore, the change in the parameter $\varepsilon_{\star{\rm dk}}$ triggers these bifurcations. As shown by Figure 2A, the critical value of $\varepsilon_{\star{\rm dk}}$, under $\psi_{\star 0} = 80°$, for the bifurcations $B_1$ and $B_2$ are approximately $0.55$ and $3.3$, respectively. It is certain that these critical values depend on $\psi_{\star 0}$ and $\varepsilon_{\star{\rm c}}$, which will be studied in our future work.

The evolution in the system's phase portrait, especially the occurrence of the bifurcations $B_1$ and $B_2$, qualitatively changes the evolutionary pathways of planets. Considering a planet initially embedded within a misaligned protoplanetary disk, the sky-blue dashed curve in the upper panel of Figure 2A represents the equilibrium path that the planet is likely to track. Before $B_2$, the orbit will consistently align with the stable equilibrium $L_3$. However, the sudden disappearance of $L_3$ at $B_2$ will lead to an instant transition in the orbit's libration. This non-adiabatic transition is critical for the subsequent dynamics, as we will show in the following sections.

## 2.3 Mutual Inclination Excitations Induced by the Saddle-Center Bifurcation

In this section, we discuss how the equilibria and their bifurcations, revealed in Sect. 2.2, influence the planetary evolution. Figure 3

presents several representative examples of dynamical evolution of our fiducial system. Three cases of $\psi_{\star 0}$ are illustrated, with each case representing a distinct type of orbit evolution.

*Type 1*. For $\psi_{\star 0} = 70^\circ$, the planet initially coincides with the equilibrium $L_3$, which approximately aligns with the orientation of the disk. When the equilibrium $L_3$ merges with the saddle point $P_5$ at the bifurcation $B_2$, there is an instantaneous transition of the equilibrium axis, about which the planet librates, from alignment with the disk to near alignment with the stellar spin axis. This non-adiabatic transition from $L_3$ to $P_5$ results in a large mutual inclination between the inner and outer orbits. Subsequently, as the stellar oblateness decays, the inner orbit gradually shifts to librate about the norm of the outer orbit. In this type of evolution, the planet follows the evolution of the equilibrium path indicated by the sky-blue dashed curve in the top panel.

*Type 2*. For $\psi_{\star 0} = 80^\circ$, the orbital evolution experiences several non-trivial processes that can be divided into several stages, as depicted in the middle of Figure 3. The dynamical behaviors are similar to those observed in the case of $\psi_{\star 0} = 70^\circ$ until the orbit is captured by the orthogonal equilibrium $L_1$ at about $3.0\times10^6$ years. Prior to this capture, the orbit librates about the equilibrium $P_5$ with a large amplitude, where $\Delta h$ circulates from $0$ to $2\pi$ (this stage referred as state I). Following the capture, the orbit begins to librate about the orthogonal equilibrium $L_1$, as indicated by $\Delta h$ librating about $-\pi/2$ (referred as state II). Finally, as the stellar oblateness decays, the orbit is recaptured by the coplanar equilibrium that is closely aligned with the norm of the outer orbit $\boldsymbol{l}_c$, where $\Delta h$ librates about $0$ (referred as state III-A). This type of evolution results in a final large mutual planetary inclination with

$i_{\mathrm{mut}} < 90°$.

*Type 3*. For $\psi_{\star 0} = 82°$, it is mostly similar to the $\psi_{\star 0} = 80°$ case, where the orbit is also captured by the orthogonal equilibrium $L_1$. However, during its subsequent transition from the orthogonal to the coplanar equilibrium, the orbit is captured by the coplanar equilibrium that is in the anti-direction of the outer orbit's angular momentum, i.e, $-\boldsymbol{l}_{\mathrm{c}}$. This scenario, contrasting with the $\psi_{\star 0} = 80°$ case, is characterized by $\Delta h$ librates about $\pi$ (referred as state III-B). Consequently, the orbit ultimately adopts a retrograde configuration with $i_{\mathrm{mut}} > 90°$.

The dynamical nature of the planetary evolution patterns in *Types 2* and *3* is essentially identical, due to the symmetry of the phase space about $\boldsymbol{l}_{\mathrm{b}}$ and $-\boldsymbol{l}_{\mathrm{b}}$, as indicated by motion equations (4) and (5). This symmetry implies that the specific orientation of the orbit's libration during state II, whether around $\hat{\boldsymbol{s}}_{\star} \times \hat{\boldsymbol{l}}_{\mathrm{dk}}$ or its negative, does not qualitatively influence the outcome of the transition to state III. After state II, the orbit randomly chooses to librate about $\boldsymbol{l}_{\mathrm{c}}$ or $-\boldsymbol{l}_{\mathrm{c}}$, resulting in a final either prograde or retrograde orbit configuration.

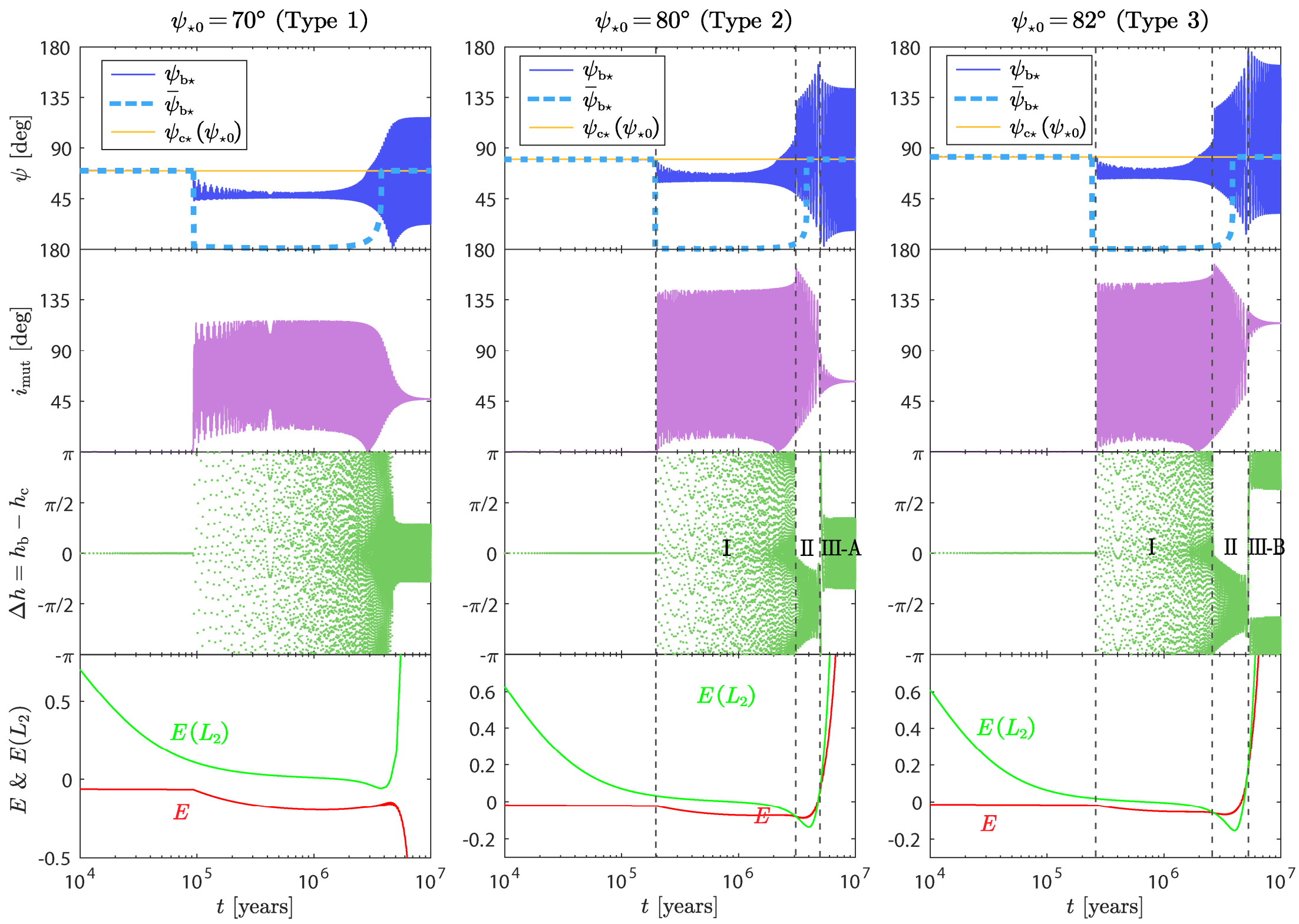

**Figure 3.** Representative examples of orbital evolution for our fiducial system. Three cases of primordial obliquity, $\psi_{\star 0} = 75°$ (left), $81°$ (middle), and $82°$ (right), are considered. For each case, the panels from the top to bottom display the temporal evolution of obliquity, mutual planetary inclination $i_{\rm mut}$, differences in the ascending node $\Delta h = h_{\rm b} - h_{\rm c}$, and the normalized energy integral $E$ and $E(L_2)$. The sky-blue dashed curve in the top panel corresponds to the same equilibrium path as depicted in Figure 2A.

Consequently, we have revealed two critical dynamical mechanisms that substantially influence the system's evolution: (1) the saddle-center bifurcation $B_2$ triggers an excitation of large mutual planetary inclinations, and (2) capture by the orthogonal equilibrium $L_1$ sets the stage for the formation of a final retrograde inner-outer orbit configuration.

## 2.4 Capture by the Orthogonal Equilibrium

In this section, we examine the dynamics of the capture by the orthogonal equilibrium during the orbital evolution. As shown by Figure 3, the capture occurs long after the bifurcation $B_2$, i.e., during the third phase identified in Figure 2. The corresponding phase portrait is illustrated in panel (c) of Figure 2B, characterized by the stable equilibria $L_1$ and $P_5$ with the separatrix emerging from $L_2$.

For the secular model outlined in Appendix (A), if we ignore the dissipation of the disk and decay of the stellar oblateness, the secular potential (or, the Hamiltonian of the system), given by equation (A1), remains constant. Alternatively, the dimensionless energy integral

$$E = -\frac{1}{2}\left(\hat{\boldsymbol{s}}_\star \cdot \hat{\boldsymbol{l}}_{\mathrm{b}}\right)^2 - \frac{1/\varepsilon_{\star\mathrm{c}}}{2}\left(\hat{\boldsymbol{l}}_{\mathrm{b}} \cdot \hat{\boldsymbol{l}}_{\mathrm{c}}\right)^2 + 1/\varepsilon_{\star\mathrm{dk}}\sqrt{1-\left(\hat{\boldsymbol{l}}_{\mathrm{dk}} \cdot \hat{\boldsymbol{l}}_{\mathrm{b}}\right)^2}\left[1-\frac{\pi}{8}\sqrt{1-\left(\hat{\boldsymbol{l}}_{\mathrm{dk}} \cdot \hat{\boldsymbol{l}}_{\mathrm{b}}\right)^2}\right] \tag{10}$$

is conserved. Furthermore, by incorporating the cconstant, $\left|\boldsymbol{l}_{\mathrm{b}}\right| = 1$, the system achieves integrability, allowing the secular equations of motion to be analytically solved.

For the conservative system, the stable equilibria occur at the local minima (or maxima) of the energy function. In our system, the equilibrium $L_1$ has the largest allowable energy $E(L_1)$, while the equilibrium $P_5$ has the minimum energy $E(P_5)$. The trajectory passing through the equilibrium $L_2$ with the energy $E(L_2)$ separates the stable libration region around $L_1$ from that around $P_5$. As a result, if $E(P_5) \le E < E(L_1)$, the orbit librates around the coplanar equilibrium $P_5$; while if $E(L_2) \le E < E(L_1)$, the orbit librates around the orthogonal equilibrium $L_1$.

Including the dissipation of the disk and the decay of the stellar oblateness introduces long-term changes in the energy function, rendering the system non-integrable. However, $\Sigma$ and $J_2$ vary over timescales significantly longer than the planet's precession period, and thus can be regarded as slowly varying parameters. Consequently, the energy function is effectively conserved during arbitrarily short periods. The libration pattern can be determined by the relationship between the instantaneous energy integral $E$ and $E(L_2)$. As shown in the bottom panels of Figure 3, a transition between libration regions occurs when these two energy curves cross each other.

We have identified three types of orbital evolution patterns in Figure 3. In addition, a fourth, dynamical trivial type exists, where all planets maintain the disk's orientation and remain coplanar. This occurs when stellar oblateness is insufficient to induce the bifurcation $B_2$. Figure 4 illustrates the domains of these four types in the $(J_{2,0},\ \psi_{\star 0})$ space. First, systems with primordial obliquities far from a polar set-up, such as $\psi_{\star 0} < 78°$ for our fiducial system, experience *Type 1* orbital precession. Second, a weak stellar oblateness generally results in *Type 4* orbital behaviors. Moreover, as the obliquity approaches $90°$, the strength of the stellar oblateness weakens due to the term $(\hat{\boldsymbol{l}}_b \cdot \hat{\boldsymbol{s}}_\star)^2$ in its potential, thus requiring a larger critical $J_{2,0}$ to trigger the bifurcation $B_2$. Finally, systems with near-polar obliquity and sufficiently strong stellar oblateness tend to exhibit orbital evolution patterns corresponding to *Type 2* or *3*.

The final mutual planetary inclinations, color-coded in the figure, show that near-polar primordial obliquity can result in nearly perpendicular orbit architecture. As predicted, within the domains of *Types 2 and 3*, the orbits probabilistically adopt one of the two types. Notably, *Type 3* evolution leads to a retrograde configuration between inner and outer orbits.

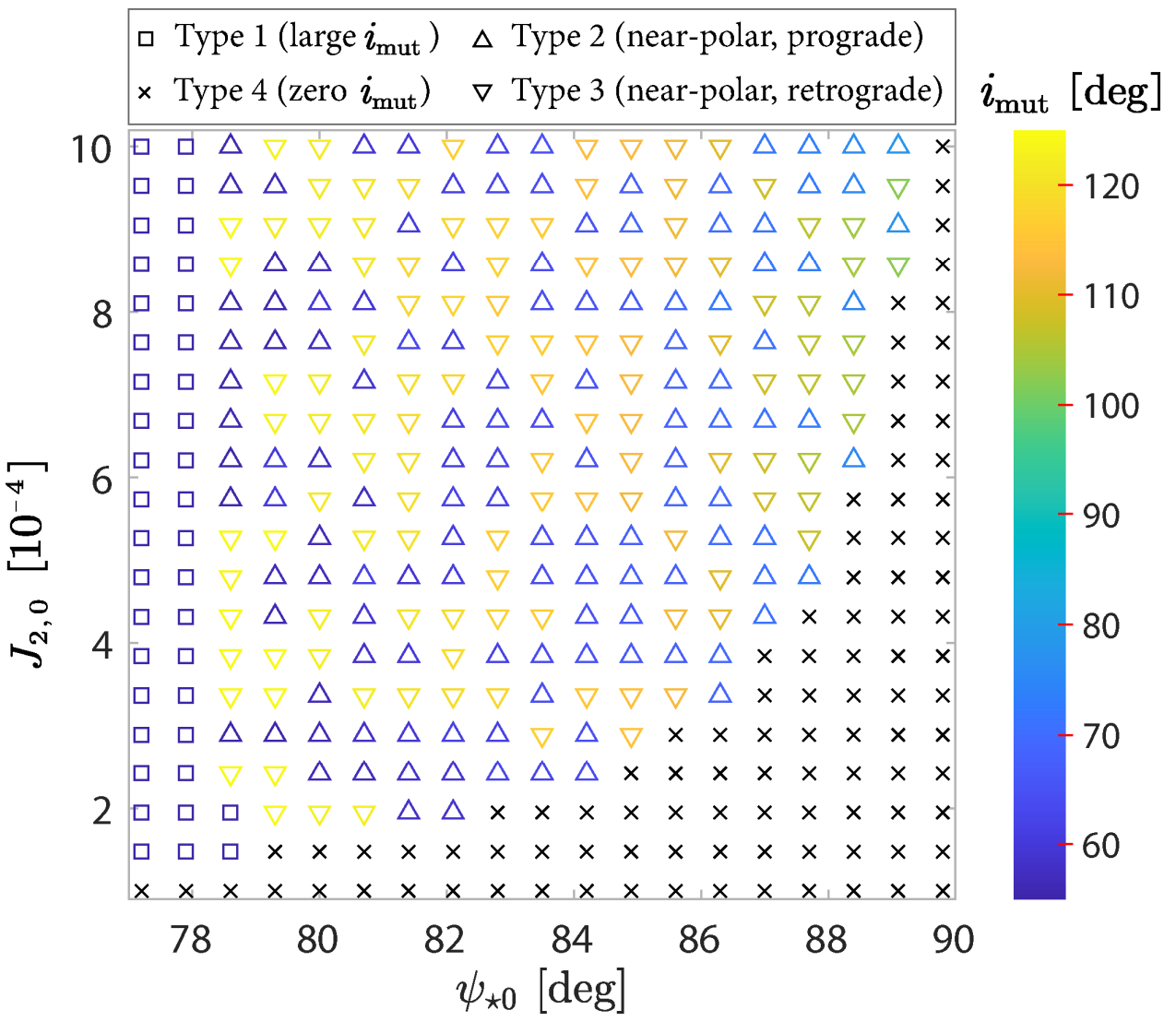

**Figure 4.** Domains of different orbit evolution types in the $(J_{2,0},\ \psi_{\star 0})$ parameter space. We vary $J_{2,0}$ from $1\times10^{-4}$ to $1\times10^{-3}$ and $\psi_{\star 0}$ from $77°$ to $90°$. Four orbit evolution types are distinguished using specific markers, and the final mutual inclination between the inner and outer orbits is color-coded. Except $J_{2,0}$ and $\psi_{\star 0}$, all other simulation parameters are those listed in Table 1 for the fiducial system. Note that systems with moderately lower $\psi_{\star 0}$ than depicted (e.g., $\psi_{\star 0} < 78°$) tend to experience *Type 1* evolution, while the weaker stellar oblateness (e.g., $J_{2,0} < 10^{-4}$) generally results in *Type 4* evolution. Results for $\psi_{\star 0} > 90°$ are expected to be similar due to the symmetry around $\psi_{\star 0} = 90°$.

## 3. Eccentricity Stability

The above analysis assumes all planets to be in circular orbits. However, these orbits may be subjected to eccentricity instability when large mutual inclinations between planets have been developed, which could undermine the validity of the previous analysis. To address this, in this section, we explore the eccentricity stability during the evolution by including general relativity.

The non-trivial orbit evolution patterns, namely *Types 1, 2* and *3*, require a relatively strong stellar oblateness to allow for a transition of the equilibrium from initially being aligned with the disk to near alignment with the stellar spin, which is attributed to the saddle-center bifurcation $B_2$. Essentially, it requires the inclined oblateness to be sufficiently strong to counterbalance the combined gravitational forces of the disk and outer planets. As illustrated in Figure 2, two parameters, $\varepsilon_{\star\mathrm{dk}}$ and $\varepsilon_{\star\mathrm{c}}$, are defined to quantify this condition. However, given the near alignment of the disk and the outer planet during the evolution and the rapid photoevaporation of the inner disk, these two parameters can be consolidated into a single parameter, which is more convenient to characterize the eccentricity stability throughout the evolutionary process, defined by

$$\epsilon_{\mathrm{obl}} = \xi \frac{\Omega_{\mathrm{b}\star}}{\Omega_{\mathrm{b,dk}} + \Omega_{\mathrm{b,c}}} \tag{11}$$

where $\xi \sim 1$ is a modify factor. This factor is defined because when $\psi_{\star 0}$ is large, the actual $\epsilon_{\mathrm{obl}}$ is smaller due to the term $\cos\psi_{\star 0}$ and thus $\xi < 1$. Empirically, we adopt $\xi = 2/3$ in our later calculations.

In addition, we define a dimensionless parameter $\epsilon_{\mathrm{GR}}$ to describe the relative strength of the general relativity,

$$\epsilon_{\mathrm{GR}} = \xi \frac{\omega_{\mathrm{b,GR}}}{\Omega_{\mathrm{b,dk}} + \Omega_{\mathrm{b,c}}} \tag{12}$$

Here, $\omega_{\mathrm{b,GR}} = \phi_{\mathrm{b,GR}} / L_{\mathrm{b}}$, where $\phi_{\mathrm{b,GR}}$ is given by equation (B2).

Consequently, the eccentricity stability can be fully characterized by the two parameters, $\epsilon_{\mathrm{obl}}$ and $\epsilon_{\mathrm{GR}}$, for a given $\psi_{\star 0}$. Figure 5 illustrates the eccentricity excitation as functions of these two parameters. As observed, without GR, orbits with $\epsilon_{\mathrm{obl}} \sim 1$ undergo dramatic chaotic excitations. Moreover, even if the initial $\epsilon_{\mathrm{obl}}$ exceeds the upper limit of the excitation range, it decreases as stellar oblateness decays, and will eventually cross this unstable range and lead to strong instability.

The eccentricity instability discussed here is attributed to the chaotic behavior resulting from the overlap of several secular resonances, which is induced by the comparable influences of stellar oblateness and the outer perturbing body. This type of eccentricity excitation can be triggered when $\psi_{\star 0} \gtrsim 45^{\circ}$ for $\epsilon_{\mathrm{obl}} \sim 1$ (Yokoyama 1999). This process is distinct from the eccentricity instability associated with the standard coplanar Laplace equilibrium, which necessitates $\psi_{\star 0} > 68.875^{\circ}$ (Tremaine et al. 2009; Saillenfest & Lari 2021).

Fortunately, general relativity significantly suppresses the excitation of eccentricity, particularly in regions where $\epsilon_{\mathrm{obl}} \lesssim \epsilon_{\mathrm{GR}}$. However, such suppressions are not guaranteed when $\epsilon_{\mathrm{obl}} > \epsilon_{\mathrm{GR}}$, as shown in Figure 5. Stability is generally maintained for nearly all $\epsilon_{\mathrm{obl}}$ values only when $\epsilon_{\mathrm{GR}}$ is substantially large, such as $\epsilon_{\mathrm{GR}} \gtrsim 25$ as indicated by the last panel of the figure. Furthermore, we observe that new regions of excitation may emerge when $\epsilon_{\mathrm{obl}} > \epsilon_{\mathrm{GR}}$, although the extent of such excitations decreases with $\epsilon_{\mathrm{GR}}$. This phenomenon may attribute to that the inclusion of GR has modified the centers of the secular resonances associated with the chaotic eccentricity excitation, thereby shifting the locations where excitation occurs. A detailed exploration of these dynamics will be the focus of our future studies.

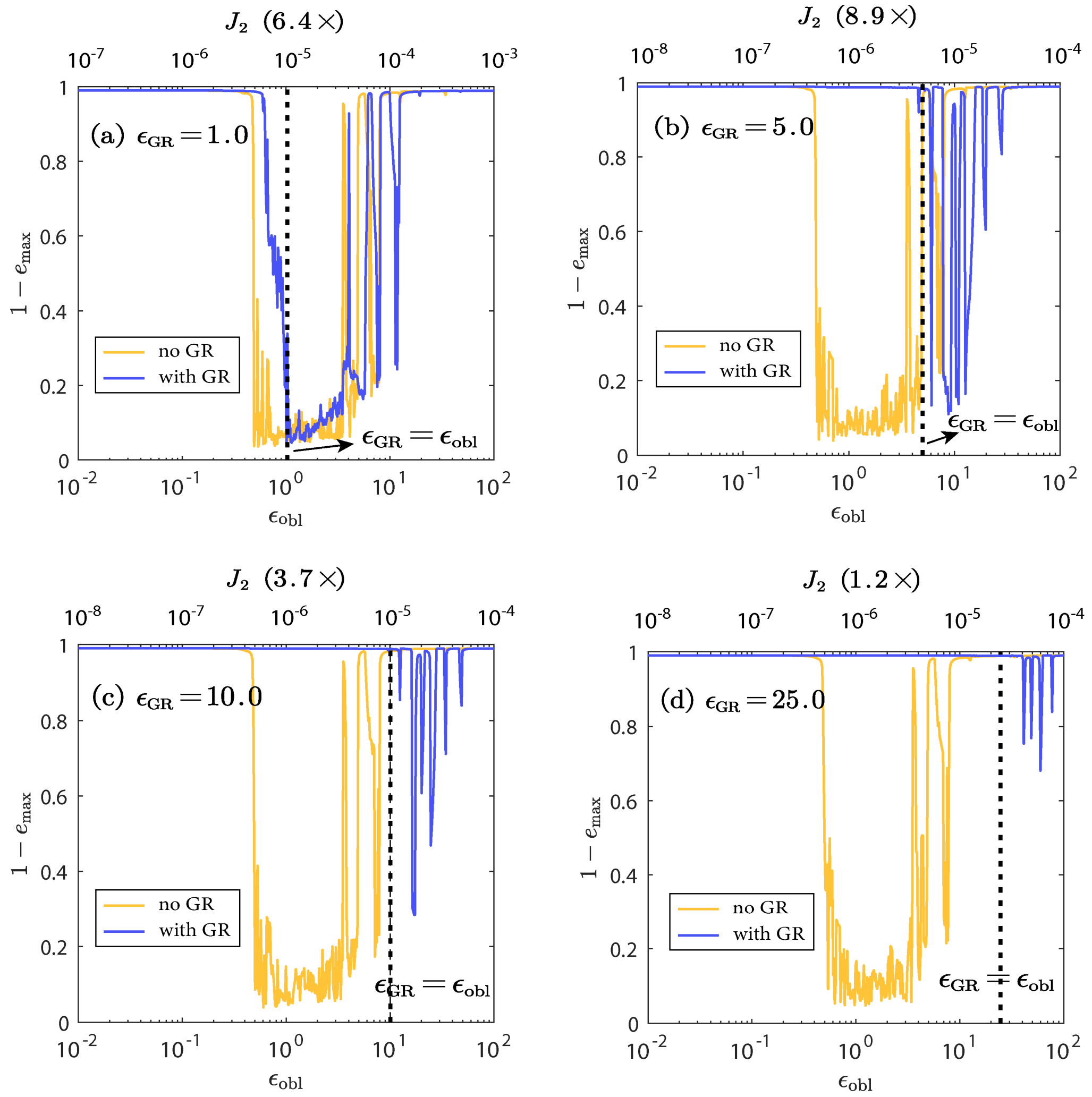

**Figure 5.** Eccentricity excitation as functions of $\epsilon_{obl}$ and $\epsilon_{GR}$. In this figure, we use the full secular model elaborated in appendix B, but without consideration of the disk potential and the decay of stellar oblateness. The initial obliquity is set as $\psi_{\star 0}=80^{\circ}$. Four cases of $\epsilon_{GR}$, (a) $\epsilon_{GR}=1.0$, (b) 5.0, (c) 10.0, and (d) 25.0, are considered. The different $\epsilon_{GR}$ values are achieved by changing the semi-major axis of the inner planet $a_b$. We vary the stellar oblateness $J_2$ in a wide range (as indicated in the top x-axis) to allow $\epsilon_{obl}$ to change from $10^{-2}$ to $10^{2}$. The maximum eccentricity $e_{max}$ during the integration up to $5\times10^{5}$ years is recorded for $\epsilon_{obl}$. Except the parameters specified, all the other simulation parameters those listed in Table 1.

# 4. Application to HD 3167

In this section, we apply the mechanisms proposed in Sects. 2 and 3 to the HD 3167 system, by using the full secular model of this system, which does not assume the lock for the outer three planets. We will demonstrate that through the proposed dynamical processes, the unique architecture of the HD 3167 system can be approximately reproduced.

## 4.1 A Parametric Window of Stability for HD 3167

First, we examine the eccentricity stability of the innermost planet, HD 3167 b. By using the system parameters listed in Table 1, HD 3167 b's $\epsilon_{\rm GR}$ is estimated to be $7.0$ for $\xi = 1$ or $4.67$ for $\xi = 2/3$ when set $\Omega_{\rm b,dkin} = 0$. Then, according to the discussions in Sect. 3, general relativity is capable of developing a parametric space for the orbit to maintain stability.

Figure 6 examines the eccentricity excitation for the HD 3167 system. Different from Figure 5, the simulations in this figure are based on the full secular model of the HD 3167 system, without the assumption of the lock of the outer three planets. The results for $\epsilon_{\rm GR} \approx 4.67$ are similar to those for $\epsilon_{\rm GR} \approx 5.0$ presented in Figure 5.

Considering both the necessary condition for the excitation of the large mutual planetary inclination, roughly $\epsilon_{\rm obl} \gtrsim 1$, and the requirement for the eccentricity stability, $\epsilon_{\rm obl} \lesssim \epsilon_{\rm GR}$, we can determine a parametric window feasible for the *Type 2* or *3* evolution of the system, as indicated in Figure 6. This window corresponds to $J_2$ varying from $3.7\times10^{-6}$ to $1.7\times10^{-5}$. However, if the disk potential is further included, the required $J_{2,0}$ should be moderately larger.

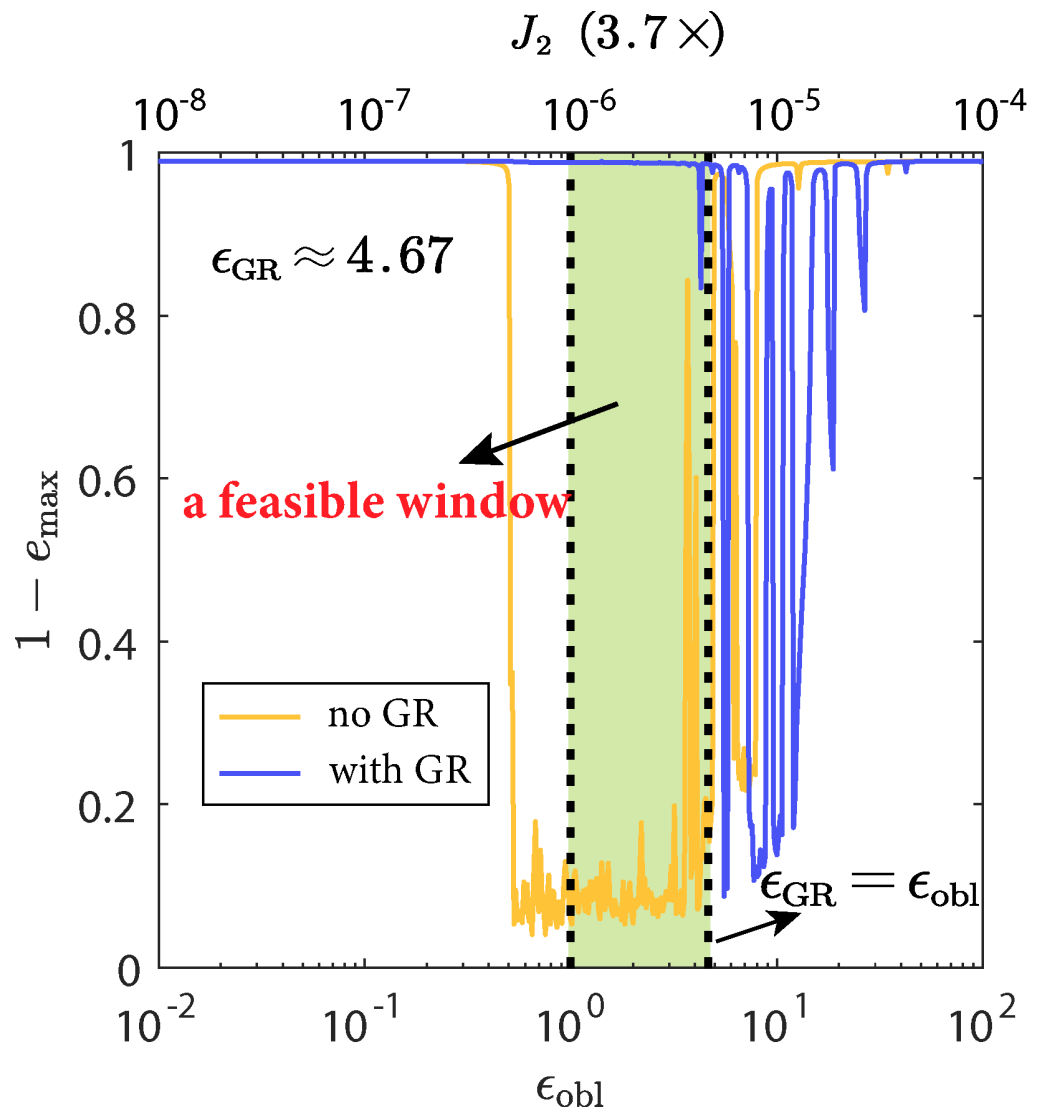

**Figure 6.** Eccentricity stability in the HD 3167 system. This figure is based on simulations using the full secular model for HD 3167 described in Appendix B, excluding effects from disk potential and stellar oblateness decay. The innermost planet's $\epsilon_{GR}$ is calculated to be about 4.67, as indicated by the right vertical dashed line. The green region, where $1 \lesssim \epsilon_{obl} \lesssim \epsilon_{GR}$, indicates conditions favorable for both large inner-outer mutual inclination and eccentricity stability. The top x-axis shows the values of $J_2$ corresponding to $\epsilon_{obl}$ values on the bottom x-axis.

## 4.2 Dynamical Evolution of the HD 3167 System

The innermost planet, HD 3167 b, with its semi-major axis $a_b = 0.018$ au, is a member of ultra-short-period planets. One of the explanations for the formation of USPs involve that they have migrated to their current locations from wider orbits through either mild low-eccentricity migrations (Pu & Lai 2019) or violent high-eccentricity processes (Petrovich et al. 2019). However, our simulations assume that HD 3167 b is formed in situ or migrated to the inner edge of the protoplanetary disk through disk migration before the rapid photoevaporation of the inner region of the disk (Serrano et al. 2022). If the migration did not complete at this stage, the upper limit of its semi-major axis can be approximately constrained by $\epsilon_{\rm GR} > 1$, resulting in $a_{\rm b} < 0.026$ au.

Figure 7 illustrates the dynamical evolution of the HD 3167 system with a near-polar primordial misalignment angle of $\psi_{\star 0} = 96°$ between the disk and the stellar spin. The innermost planet maintains stability throughout the simulation due to $\epsilon_{\rm obl} \lesssim \epsilon_{\rm GR}$, for the initial stellar oblateness chosen as $J_{2,0} = 2\times10^{-5}$. The outer three planets also exhibit stability and remain coplanar, preserving the disk's orientation with $\psi_{i\star} \approx \psi_{\star 0}$. The innermost planet undergoes type 3 evolution, resulting in a near-perpendicular, retrograde configuration between the inner and outer orbits. The resultant stellar obliquities $\psi_{\rm b\star}$ and $\psi_{\rm c\star}$, and the mutual inclination, $i_{\rm bc}$, are close to their observation constrained values, as shown in the figure. Note the planet, HD 3167 b, eventually librates around the norm of the outer orbits with a large amplitude, as the stellar oblateness sufficiently weakens. Consequently, its observed close alignment with the stellar spin, characterized by $\psi_{\star \rm b} \approx 29°$, is actually a temporary state from a long-term perspective, as also suggested by Bourrier et al. (2022).

In this study, considering the huge parameter space and other possible factors that can affect the system's subsequent dynamics, we do not seek to recover the parameters and processes that exactly reproduce the observed architecture of the HD 3167 system.

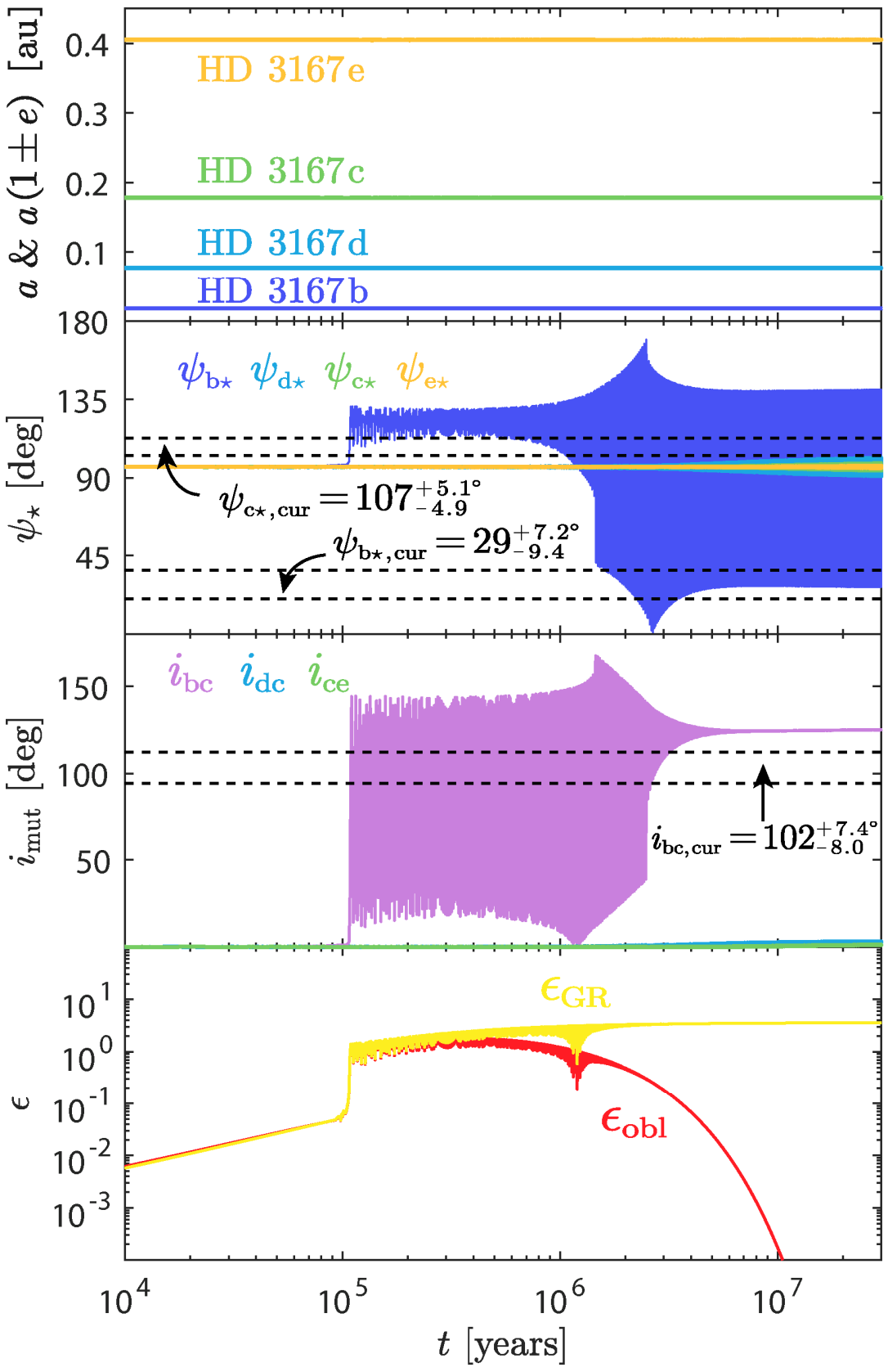

**Figure 7.** Dynamical evolution of the HD 3167 system with a near-polar misaligned protoplanetary disk. This simulation uses the full secular model for HD 3167 described in Appendix B. We adopt $\psi_{\star,0} = 96°$ and $J_{2,0} = 2\times10^{-5}$. Since HD 3167 b with $a_{\mathrm{b}} \approx 0.18$ au is extremely close to the host star, we further assume the disk within $0.05$ au dissipates with the timescale $\tau^{*}_{\mathrm{v,in}} = 10^{3}$ yr. All the other parameters are listed in Table 1. From top to bottom panels, obliquities, mutual planetary inclinations, and the dimensionless parameters $\epsilon_{\mathrm{obl}}$ and $\epsilon_{\mathrm{GR}}$, are depicted. The observation constrained values for $\psi_{\mathrm{b}\star}$, $\psi_{\mathrm{c}\star}$, and $i_{\mathrm{bc}}$, with their $1\sigma$ uncertainties, are also indicated.

## 5. Discussion

In this study, we investigated the dynamics of multi-planet systems, driven by the combined effects of a misaligned protoplanetary disk and stellar oblateness. For the first time, we demonstrated that a near-polar disk can produce a retrograde, near-perpendicular orbit configuration, similar to the one observed in HD 3167. One of our testable predictions for such systems is that the outer planets will maintain the disk's orientation with a high obliquity, while the obliquity of the innermost planet could exhibit a broad distribution.

## 5.1 Robustness of the Saddle-Center Bifurcation and the Capture by the Orthogonal Equilibrium

As discussed in Sect. 2, the saddle-center bifurcation $B_2$ occurring during the photoevaporation of the disk is essential for generating large mutual planetary inclinations. This bifurcation is hardly dependent on the dissipation timescale of the inner disk $\tau_{\mathrm{v,in}}$, which is estimated to be about $10^2$–$10^5$ years(Owen 2016; Zanazzi & Lai 2018). This mechanism contrasts qualitatively with the one that produces moderate mutual inclinations through differential precessions among planetary orbits driven by stellar oblateness, a process that is highly sensitive to $\tau_{\mathrm{v,in}}$ (Li et al. 2020; Spalding & Batygin 2016; Spalding & Millholland 2020).

The capture by the orthogonal equilibrium is primarily induced by the decay of the stellar oblateness, which is essential for the *types 2* and *3* evolution. The decay of the pre-main-sequence stellar oblateness is attributed to the stellar contraction, spin down, and evolution of the tidal Love number. The evolution of the stellar radius can be approximately described by (Batygin & Adams 2013)

$$R_\star = R_{\star,0}\left(1+t/\tau_{R\star}\right)^{-1/3} \tag{13}$$

where $\tau_{R\star} \sim 1$ Myr. Consequently, given the oblateness formulation in equation (A16), the term $J_2 R_\star^2$ evolves according to

$$J_2 R_\star^2 = J_{2,0} R_{\star,0}^2 \left(1 + t/\tau_{R\star}\right)^{-\gamma} \tag{14}$$

where $\gamma = 5/3$. Moreover, if the stellar spin down is further considered, $\gamma > 5/3$ can be expected (Millholland & Laughlin 2019).

Through systematic exploration of the parameter space $(\gamma,\ \tau_{\mathrm{v,in}})$ by varying $\gamma$ from $5/3$ to $2.5$ and $\tau_{\mathrm{v,in}}$ from $10^2$ to $10^5$ years, we find that as long as the initial stellar oblateness is sufficiently strong, the bifurcation and the capture by the orthogonal equilibrium occur, thereby demonstrating the robustness of the proposed mechanism.

## 5.2 Prevalence of the Proposed Mechanism

The efficacy of the proposed mechanism hinges primarily on two factors: (1) the prevalence of primordial disk misalignments, and (2) the satisfaction of the stability requirements.

Regarding the first factor, highly misaligned protoplanetary disk can theoretically arise from magnetic warping and gravitational torque from binary companions. Notably, the later mechanism tends to frequently produce high and even polar disk misalignments. When combined with high-eccentricity migration, this can lead to a subset population of perpendicular hot Jupiters, similar to the recently observed abundance of perpendicular planets among hot Jupiter systems (Vick et al. 2023). Additionally, polar circumbinary disks have been recently observed and confirmed (Kennedy et al. 2019). Planets formed in these disks may undergo similar dynamical processes to those we have considered here (Farago & Laskar 2010).

As for the second factor, in addition to the stability of the innermost planet, the outer planets are also subjected to eccentricity instability. For example, compact multi-planet systems from the Kepler sample often struggle to maintain stability in the presence of high primordial obliquities (Spalding et al. 2018; Schultz et al. 2021). However, there are two scenarios in which planetary systems can remain stable. The first involves systems like HD 3167, where the outer planets are weakly coupled to the innermost planet. The second scenario involves the outer planet being significantly more massive than the inner planet (i.e., a giant planet), which allows the outer planet to maintain stability. This latter scenario may be quite common, given that approximately 40% of close-in, small planet have been observed to have an outer giant planet companion, based on radial velocity survey (Rosenthal et al. 2022). However, another recent study reveals a lower fraction of about 9% (Bonomo et al. 2023).

## 5.3 Applications to Other Observed Systems

To begin with, we compare our mechanism with the disk-dispersal driven resonance proposed by Petrovich et al. (2020). They assume an initial alignment between the disk and the stellar equator, which leads to their prediction of a low obliquity for the outer planet due to the conservation of the total angular momentum deficit. Therefore, this mechanism can hardly be applied to systems like HD 3167 where the outer planets are in highly obliqued orbits. Moreover, their model generally does not anticipate a retrograde configuration between the inner and outer planets. In contrast, our mechanism anticipates a high obliquity for the outer planets as a result of the primordial disk misalignment, while the obliquity of the inner planet can span a broad range. Most importantly, our model predicts the occurrence of a retrograde, perpendicular orbit configuration between the inner and outer planets.

Next, we examine several sub-Neptunes that have been observed with high obliquities, as catalogued at TEPCat (Southworth 2011). (Planets with high obliquities but without detected outer planets are not included).

1. HAT-P-11 has a near-polar inner planet ($a_b \simeq 0.052$ au) with an obliquity of $\psi_\star = 104.9^{+8.6°}_{-9.1}$ (Bourrier et al. 2023), an outer giant planet companion ($m_c \simeq 2.3 m_J$ and $a_c \simeq 4.13$), and a stellar mass of $m_\star \simeq 0.809 m_\odot$, resulting in $\epsilon_{GR} \sim 73 \gg 1$. The excitation of the mutual inclination requires $\epsilon_{obl} > 1$, corresponding to a lower limit for the stellar oblateness of $J_2 \gtrsim 10^{-6}$. Consequently, this system is a potential candidate for our proposed mechanism.
2. WASP-107 has a similar architecture to HAT-P-11: an inner planet ($a_b \simeq 0.055$ au) with $\psi_\star = 103.5^{+1.79°}_{-1.8}$ and an outer giant planet ($m_c \simeq 0.36 m_J$ and $a_c \simeq 1.82$ au). The inner planet's $\epsilon_{GR}$ is estimated to be about $23$ satisfying the stability requirement. For this system, the excitation of the mutual inclination necessitates $J_{2,0} \gtrsim 3\times10^{-6}$.
3. $\pi$ Mensae has been observed with three planets. The feasible parametric window for the proposed mechanism to operate is estimated to be quite small; therefore, the large mutual inclination between planets b and c ($49° < i_{bc} < 131°$) is unlikely to be produced by primordial disk misalignments.
4. 55 Cancri has an USP with $\psi_\star = 23^{+14°}_{-12}$ (Zhao et al. 2022), and four outer planets whose obliquities remain unclear but are potentially moderately inclined relative to the innermost planet. By using the system parameters, the USP's $\epsilon_{GR}$ is computed to be approximately 0.8 (assuming coplanarity of the outer planets), indicating that a moderate primordial disk misalignment could be possible, but a highly tilted disk is unlikely.

Here, we only examine the fulfillment of stability conditions and the lower requirement of the stellar oblateness for excitation of the mutual inclination. Identifying the actual causes of their high obliquities needs additional observational constraints. For instance, if outer massive planets are further observed to be located in highly oblique or even polar orbits, it could provide significant evidence supporting this process.

## 6. Conclusions

We have investigated the dynamics of close-in planets within a multi-planet system, driven by the combined effects of a misaligned protoplanetary disk and stellar oblateness. Through the newly proposed mechanism, the retrograde, near-perpendicular configuration observed within the HD 3167 system can be explained, based on the assumption that the USP, HD 3167 b, is formed in situ or migrated to the inner edge of the disk through disk migration prior to the rapid photoevaporation of the disk.

The key mechanism we have discovered is that a saddle-center bifurcation, occurring during the photoevaporation of the disk, can induce a large mutual planetary inclination. This bifurcation triggers an instant, non-adiabatic transition in the planet's orbit libration. Then, in scenarios of near-polar primordial misalignment, the capture by the orthogonal equilibrium provides a pathway for the development of retrograde inner-outer orbit configurations. Additionally, general relativity plays a crucial role by creating a parametric window that allows the planet to maintain stability during these complex dynamical evolutions.

We predict that systems undergoing this dynamical process should exhibit specific, testable characteristics. For instance, such systems should meet stability requirements, and the outer planets likely exhibit high stellar obliquities as they tend to retain the disk's orientation, while the obliquity of the innermost planet may vary widely.

## Acknowledgments

We gratefully acknowledge the referee for providing a constructive review report of our work. We thank Will M. Farr for his contribution to the publicly available Rings code (at https://github.com/farr/Rings). This work has been supported by the National Natural Science Foundation of China under grant 11872007, and the Fundamental Research Funds for the Central Universities.

## Appendix A

## Equations of motion of the Circular Problem

Considering a two-planet system (with planetary masses $m_{\mathrm{b}}$ and $m_{\mathrm{c}}$, and semi-major axes $a_{\mathrm{b}}$ and $a_{\mathrm{c}}$ satisfying $a_{\mathrm{b}} \ll a_{\mathrm{c}}$) initially embedded within a misaligned protoplanetary disk, the forces and effects that govern the secular dynamics of the planets involve the rotation-induced oblateness (oriented along $\hat{\boldsymbol{s}}_{\star}$), gravity of the disk (oriented along $\hat{\boldsymbol{l}}_{\mathrm{d}}$), and mutual planetary interactions. We first develop a simplified model based on the assumptions that both the planetary orbits are circular ($e(s)=0$), and the outer orbit is fixed due to $L_b \ll L_c$, where $L_b$ and $L_c$ are the magnitudes of angular momenta of the inner and outer orbits, respectively. Then, the secular potential that describes the motion of the inner planet, in terms of the normalized angular momentum vector $\boldsymbol{l} = \sqrt{1-e^2}\hat{\boldsymbol{l}}$ and eccentricity vector $\boldsymbol{e} = e\hat{\boldsymbol{e}}$, is given by(Tremaine & Yavetz 2014; Fu et al. 2023)

$$\begin{aligned}\Phi &= \Phi_{b\star} + \Phi_{b,c} + \Phi_{dk,b} \\ &= -\frac{\phi_{b\star}}{2}\left(\hat{\boldsymbol{s}}_\star \cdot \hat{\boldsymbol{l}}_b\right)^2 - \frac{\phi_{b,c}}{2}\left(\hat{\boldsymbol{l}}_b \cdot \hat{\boldsymbol{l}}_c\right)^2 \\ &+ \phi_{dk,b}\sqrt{1-\left(\hat{\boldsymbol{l}}_{dk} \cdot \hat{\boldsymbol{l}}_b\right)^2}\left[1 - \frac{\pi}{8}\sqrt{1-\left(\hat{\boldsymbol{l}}_{dk} \cdot \hat{\boldsymbol{l}}_b\right)^2}\right]\end{aligned} \quad . \tag{A1}$$

Here,

$$\phi_{b\star} = \frac{3Gm_\star m_b J_2 R_\star^2}{2a_b^3} \tag{A2}$$

$$\phi_{b,c} = \frac{3Gm_b m_c a_b^2}{4a_c^3\left(1-e_c^2\right)^{3/2}} \tag{A3}$$

$$\phi_{dk,b} = 2\pi G m_b \Sigma_0 a_0 \tag{A4}$$

where $G$ is the universal gravitational constant, $m_\star$ and $R_\star$ are the mass and radius of the central star, and $\Sigma_0$ is the disk surface density at radius $a_0$. The secular disk potential $\Phi_{dk,b}$, derived from the direct potential of an infinite Mestel disk (Thommes et al. 2008; Schulz 2012), will be detailed below. Here, $\Phi_{dk,b}$ represents $\Phi_{dkin,b}$ given by equation (A11).

In the above equations, the interaction potential $\Phi_{b,c}$ between the two planets is described by the quadrupole-level approximation of the expansion in the ratio of their semimajor axes $a_b/a_c$. If the hierarchy ($a_b \ll a_c$) is not well satisfied, $\Phi_{b,c}$ can be described by using Laplace-Lagrange theory based on the assumptions of small eccentricities and inclinations, with $\phi_{b,c}$ taking the form

$$\phi_{\mathrm{b,c}} = \frac{G m_{\mathrm{b}} m_{\mathrm{c}} a_{\mathrm{b}}}{4 a_c^2} b_{3/2}^{(1)}(\alpha) \tag{A5}$$

where $\alpha = a_{\mathrm{b}}/a_{\mathrm{c}}$ and $b_{3/2}^{(1)}(\alpha)$ is a Laplace coefficient, defined by

$$b_{3/2}^{(1)}(\alpha) = \frac{1}{\pi}\int_0^{2\pi} \frac{\cos\theta}{\left(1 - 2\alpha\cos\theta + \alpha^2\right)^{3/2}} \mathrm{d}\theta \tag{A6}$$

However, even under the circular assumptions, this approximation lose accuracy if the large mutual planetary inclinations are established.

The disk potential $\Phi_{\mathrm{dk,\,b}}$ depends on the evolution of the disk. The disk surface density ($\Sigma$) is dramatically influenced by the combined effects of photoevaporation and viscous accretion (Alexander et al. 2014; Owen 2016). Initially, the depletion of the disk mass is dominated by the viscous accretion. When reaches a certain characteristic time $\tau_{\mathrm{w}}$, the photoevaporation driven mass loss rate becomes comparable to that by the viscous accretion in the inner regions. After the characteristic time $\tau_{\mathrm{w}}$, photoevaporation starves the inner region of the disk ($r < r_{\mathrm{c}}$ ~ a few au) from resupply by the outer disk's viscous evolution, leading to a rapid dissipation of the inner disk with the viscous time $\tau_{\mathrm{v,\,in}}$, while the outer disk ($r > r_{\mathrm{c}}$) continues to evolve over the viscous time $\tau_{\mathrm{v,\,out}} \gg \tau_{\mathrm{v,\,in}}$.

It is reasonable to believe that before $\tau_{\mathrm{w}}$, all the planets are well-aligned with the disk due to its dominant effects, and thereby the planets' evolution is trivial. Therefore, we focus on the dynamics after $\tau_{\mathrm{w}}$. The evolution of the disk surface density that captures the main effect of photoevaporation can be parameterized as (assuming both the inner and outer disks follow a Mestel disk density profile)

$$\Sigma(r,t) = \begin{cases} \Sigma_{\mathrm{c}}(t)\left(r_{\mathrm{c}}/r\right), r_{\mathrm{in}} \leqslant r \leqslant r_{\mathrm{c}} \\ \Sigma_{\mathrm{out}}(t)\left(r_{\mathrm{out}}/r\right), r_{\mathrm{c}} < r \leq r_{\mathrm{out}} \end{cases}, \tag{A7}$$

where $r_{\mathrm{in}}$ and $r_{\mathrm{out}}$ are the inner and outer truncation radii of the disk,

and

$$\Sigma_{\rm c}(t) = \Sigma_{\rm c}(0) \big/ \left(1 + t/\tau_{\rm v,in}\right)$$
$$\Sigma_{\rm out}(t) = \Sigma_{\rm out}(0) \big/ \left(1 + t/\tau_{\rm v,out}\right) \tag{A8}$$

The disk mass is then (assuming $r_{\rm in} \ll r_{\rm out}$)

$$m_{\rm d}(t) \simeq 2\pi\Sigma_{\rm out}(t) r_{\rm out}^2 \tag{A9}$$

The direct potential of the inner Mestel disk ($r < r_{\rm c}$) for a close-in planet (planet $i$ with semimajor axis $a_i \ll r_{\rm c}$ in the full physical space is approximated by (Thommes et al. 2008; Schulz 2012; Spalding & Millholland 2020)

$$\Phi_{{\rm dkin},i} = \pi^2 G m_i \Sigma_a(t) a \ln\left(\left|\hat{\boldsymbol{l}}_{\rm dk} \cdot \boldsymbol{r}_i\right| + r_i\right) \tag{A10}$$

Averaging the potential over the planet's mean orbit motion, the corresponding secular pattern is given by

$$\begin{aligned} \Phi_{{\rm dkin},i} = \pi^2 G m_i \Sigma_a a \Bigg\{ & \frac{e_i^2}{4} + \frac{2}{\pi} \frac{\sqrt{1-e_i^2}}{\sqrt{1-\left(\hat{\boldsymbol{l}}_{\rm dk} \cdot \hat{\boldsymbol{l}}_i\right)^2}} \Bigg[ 1 - \left(\hat{\boldsymbol{l}}_{\rm dk} \cdot \hat{\boldsymbol{l}}_i\right)^2 \\ & + \frac{e_i^2}{2}\left(\hat{\boldsymbol{l}}_{\rm dk} \cdot \hat{\boldsymbol{e}}_i\right)^2 \Bigg] - \frac{1}{16}\Bigg[ 4 - 3e_i^2 + 6e_i^2\left(\hat{\boldsymbol{l}}_{\rm d} \cdot \hat{\boldsymbol{e}}_i\right)^2 \\ & - \left(4 - 3e_i^2\right)\left(\hat{\boldsymbol{l}}_{\rm d} \cdot \hat{\boldsymbol{l}}_i\right)^2 \Bigg] \Bigg\} + o\left(e_i^3\right) \end{aligned} \tag{A11}$$

in which we ignore terms higher than $e^2$. If substitute $e = 0$, it takes a simple form as presented in equation (A1).

In several previous studies, a secular potential of the disk assuming the planet lie close to the disk plane is adopted, which results in $\boldsymbol{l}_i$ precessing around $\hat{\boldsymbol{l}}_{\rm dk}$ at a rate (Ward 1981; Hahn 2003)

$$\Omega_{i,{\rm dkin}}^{\rm ca} \simeq \frac{\pi G \Sigma_a}{n_i a_i \beta(a_i)} \tag{A12}$$

This equation was derived under the assumption $\left|\hat{\boldsymbol{l}}_i \times \hat{\boldsymbol{l}}_{\rm dk}\right| \ll \beta(a_i) \ll 1$, where $\beta(a_i)$ is the disk aspect ratio $\beta = H/r$ ($H$ is the disk scale height) evaluated

at $r = a_i$ (Zanazzi & Lai 2018).

The full-space disk potential from equation (A11), assuming $e_i = 0$ and constraining $\hat{\boldsymbol{l}}_i \cdot \hat{\boldsymbol{l}}_{\rm dk} \ll 1$, results in

$$\Omega_{i,\rm dkin}^{\rm fs} \simeq \frac{2\pi G \Sigma_a}{n_i a_i \theta_{i,\rm dk}} \tag{A13}$$

where $\theta_{i,\rm dk}$ is the angle between $\hat{\boldsymbol{l}}_i$ and $\hat{\boldsymbol{l}}_{\rm dk}$, or the inclination of the planet relative to the disk. If we choose a softening scale $\beta(a_i)$ in equation (A13), $\Omega_{i,\rm dkin}^{\rm fs}$ becomes

$$\Omega_{i,\rm dkin}^{\rm fs} \simeq \frac{2\pi G \Sigma_a}{n_i a_i \left[\theta_{i,\rm dk} + \beta(a_i)\right]} \tag{A14}$$

If evaluated at $\theta_{i,\rm dk} = \beta(a_i)$, we can obtain $\Omega_{i,\rm dkin}^{\rm ca} = \Omega_{i,\rm dkin}^{\rm fs}$. In our numerical simulations, we also consider such a softening scale.

Considering $a_i \ll r_{\rm c}$, the secular potential of the outer disk ($r > r_{\rm out}$) for a close-in planet is given by

$$\Phi_{\mathrm{dkout},i} = \frac{3\pi G m_i a_i^2 \Sigma_{\rm out} r_{\rm out}}{8 r_{\rm c}^2 l_i^3}\left[5\left(\boldsymbol{e}_i \cdot \hat{\boldsymbol{l}}_{\rm dk}\right)^2 - \left(\boldsymbol{l}_i \cdot \hat{\boldsymbol{l}}_{\rm dk}\right)^2 + 1/3 - 2e_i^2\right] \tag{A15}$$

It has a similar expression as the potential from an outer perturbing body. In equation (A1), the secular potential of the outer disk on planet b is ignored, in which $\Phi_{\rm dk,b}$ is actually $\Phi_{\rm dkin,b}$.

In addition to the evolution of the disk, the stellar oblateness $J_2$ decays over time. The stellar rotation-induced oblateness $J_2$ can be estimated by (Sterne 1939)

$$J_2 = \frac{2}{3} k_{\rm q\star} \frac{\Omega_\star^2}{G m_\star / R_\star^3} \tag{A16}$$

where $k_{\rm q\star}$ is the apsidal motion constant and $\Omega_\star$ is the rotation velocity. The pre-main-sequence stellar oblateness is believed to be initially significant due to its expand radius $R_\star$ and short rotational period $P_\star$ (typically 3–10 days for T-Tauri stars; Bouvier et al. (2014)). But, the value of $J_2$ decays over time due to the contraction in radius, spin down, and evolution of $k_{q\star}$. In our simulations, we simply allow $J_2$ decay exponentially through (holding other parameters constant)

$$J_2(t) = J_{2,0} e^{-t/\tau_\star} \tag{A17}$$

where we adopt $\tau_\star = 1$ Myr, which roughly coincides with a Kelvin-Helmholtz timescale (Batygin & Adams 2013).

Finally, for the circular problem, we solve the motion of $\boldsymbol{l}_\mathrm{b}$ using the vectorial equations

$$\frac{\mathrm{d}\boldsymbol{l}_\mathrm{b}}{\mathrm{d}t} = -\frac{1}{L_\mathrm{b}} \boldsymbol{l}_\mathrm{b} \times \nabla_{\boldsymbol{l}_\mathrm{b}} \Phi \tag{A18}$$

where $L_\mathrm{b} = m_\mathrm{b}\sqrt{G m_\star a_\mathrm{b}}$ is the planet b's orbital angular momentum.

# Appendix B

# The full secular model

*The full secular model of the fiducial two-planet system.* The secular potential outlined in equation (A1) operates under the assumption that all planetary orbits are circular. The full secular potential considering the planetary eccentricity is given by

$$\begin{aligned} \Phi &= \Phi_{\mathrm{b}\star} + \Phi_{\mathrm{b,GR}} + \Phi_{\mathrm{b,c}} + \sum_{i=\{\mathrm{b,c}\}} \left(\Phi_{\mathrm{dkin},i} + \Phi_{\mathrm{dkout},i}\right) \\ &= \frac{\phi_{\mathrm{b}\star}}{6 l_\mathrm{b}^3}\left[1 - 3\left(\hat{\boldsymbol{s}}_\star \cdot \hat{\boldsymbol{l}}_\mathrm{b}\right)^2\right] - \frac{\phi_{\mathrm{b,GR}}}{l_\mathrm{b}} \\ &+ \frac{\phi_{\mathrm{b,c}}}{2 l_\mathrm{b}^3}\left[5\left(\boldsymbol{e}_\mathrm{b} \cdot \hat{\boldsymbol{l}}_\mathrm{c}\right)^2 - \left(\boldsymbol{l}_\mathrm{b} \cdot \hat{\boldsymbol{l}}_\mathrm{c}\right)^2 + 1/3 - 2e_\mathrm{b}^2\right] \\ &+ \sum_{i=\{\mathrm{b,c}\}} \left(\Phi_{\mathrm{dkin},i} + \Phi_{\mathrm{dkout},i}\right) \end{aligned} \tag{B1}$$

Here, $\phi_{b\star}$ and $\phi_{\mathrm{b,c}}$ are presented by equations (A2) and (A3), $\phi_{\mathrm{b,GR}}$

represents the contribution of general relativity, taking the form

$$\phi_{\mathrm{b,GR}} = \frac{3G^2 m_\star^2 m_{\mathrm{b}}}{a_{\mathrm{b}}^2 c^2} \tag{B2}$$

where $c$ is the light speed, and $\Phi_{\mathrm{dkin,\,b}}$ and $\Phi_{\mathrm{dkout,\,b}}$ are given by equations (A11) and (A15), respectively.

Then, the secular equations of motion in terms of the vectorial elements, $\boldsymbol{l}$ and $\boldsymbol{e}$, can be derived from the Lagrange planetary equations

$$\begin{aligned} \dot{\boldsymbol{l}}_i &= -\frac{1}{L_i}\left(\boldsymbol{l}_i \times \nabla_{\boldsymbol{l}_i}\Phi + \boldsymbol{e}_i \times \nabla_{\boldsymbol{e}_i}\Phi\right) \\ \dot{\boldsymbol{e}}_i &= -\frac{1}{L_i}\left(\boldsymbol{l}_i \times \nabla_{\boldsymbol{e}_i}\Phi + \boldsymbol{l}_i \times \nabla_{\boldsymbol{e}_i}\Phi\right) \end{aligned} \tag{B3}$$

The exchange of angular momenta of the disk and stellar spin is also included in our simulations, which can be described by

$$\begin{aligned} \dot{\hat{\mathbf{s}}}_\star &= -\Omega_{\star\mathrm{dk}}\left(\hat{\boldsymbol{l}}_{\mathrm{dk}} \cdot \hat{\mathbf{s}}_\star\right)\hat{\boldsymbol{l}}_{\mathrm{dk}} \times \hat{\mathbf{s}}_\star \\ \dot{\hat{\boldsymbol{l}}}_{\mathrm{dk}} &= -\Omega_{\mathrm{dk}\star}\left(\hat{\boldsymbol{l}}_{\mathrm{dk}} \cdot \hat{\mathbf{s}}_\star\right)\hat{\mathbf{s}}_\star \times \hat{\boldsymbol{l}}_{\mathrm{dk}} \end{aligned}, \tag{B4}$$

where

$$\Omega_{\star\mathrm{dk}} = \frac{k_{\mathrm{q}\star}}{k_\star}\frac{\pi R_\star^3}{m_\star}\Omega_\star\left[\frac{\Sigma_{\mathrm{c}} r_{\mathrm{c}}}{r_{\mathrm{in}}^2} + \frac{\Sigma_{\mathrm{out}} r_{\mathrm{out}}}{r_{\mathrm{c}}^2}\right] \tag{B5}$$

and $\Omega_{\mathrm{dk}\star} = \Omega_{\star\mathrm{dk}}\left(S_\star / L_{\mathrm{dk}}\right)$, in which $S_\star = k_\star m_\star R_\star^2 \Omega_\star$ and $L_{\mathrm{dk}} = 2/3\, m_{\mathrm{dk}}\sqrt{G m_\star r_{\mathrm{out}}}$ are the angular momenta of the stellar spin and disk, with $k_\star \simeq 0.2$.

*The full secular model for HD 3167*. The secular potential for the HD 3167 system that includes the similar secular effects can be given by

$$\begin{aligned} \Phi &= \Phi_{\mathrm{b}\star} + \Phi_{\mathrm{b,GR}} + \sum_{i=\{\mathrm{d,c,e}\}}\Phi_{\mathrm{b},i} + \sum_{i=\{\mathrm{b,d,c,e}\}}\left(\Phi_{\mathrm{dkout},i} + \Phi_{\mathrm{dkin},i}\right) + \sum_{i,j=\{\mathrm{d,c,e}\},\, i\neq j}\Phi_{i,j} \\ &= \frac{\phi_{\mathrm{b}\star}}{6l_{\mathrm{b}}^3}\left[1 - 3\left(\hat{\boldsymbol{s}}_\star \cdot \hat{\boldsymbol{l}}_{\mathrm{b}}\right)^2\right] - \frac{\phi_{\mathrm{GR,b}}}{l_{\mathrm{b}}} \\ &+ \sum_{i=\{\mathrm{d,c,e}\}}\frac{\phi_{\mathrm{b},i}}{2l_i^3}\left[5\left(\boldsymbol{e}_{\mathrm{b}} \cdot \hat{\boldsymbol{l}}_i\right)^2 - \left(\mathbf{l}_{\mathrm{b}} \cdot \hat{\boldsymbol{l}}_i\right)^2 + 1/3 - 2e_{\mathrm{b}}^2\right] \\ &+ \sum_{i=\{\mathrm{b,d,c,e}\}}\left(\Phi_{\mathrm{dkout},i} + \Phi_{\mathrm{dkin},i}\right) + \sum_{i\neq j}\Phi_{i,j} \end{aligned} \tag{B6}$$

where the subscripts "b", "d", "c", and "e" represent the planets HD 3167

b-e, respectively. In this equation, the interactions between the innermost planet and the outer planets, $\Phi_{b,i}$, are described by the quadruple-level approximation based on its hierarchical configuration, while the interactions among the outer three planets, $\phi_{i,j}$, can be described by either the Laplace-Lagrange theory (equation (A5)) or the Gauss ring method. The Laplace-Lagrange theory is only valid when the outer three planets are closely aligned, and have no eccentricity excitations. The Gauss method is a numerical averaging algorithm that provides an efficient way of precisely computing the secular interactions between planets, without radius ratio, eccentricity, and inclination truncations (refer to Touma et al. (2009) for details). However, this method offers no analytical expression for $\Phi_{i,j}$. To capture the potential eccentricity and mutual inclination excitations among the outer planets, we have adopted the Gauss ring method for Figure 7, although in this case, the Laplace-Lagrange theory is also feasible.

Except for $\Phi_{i,j}$, the secular equations derived from the potentials in equation can be obtained by using equation (B3). Then, we incorporate these equations along with equation (B4) into the publicly available Gauss ring code, the full secular dynamics of the HD 3167 system can be numerically simulated.